\documentclass[aip,jcp,reprint]{revtex4-1}
\usepackage{graphicx}
 \usepackage{bm} 
\usepackage[colorlinks]{hyperref}
\usepackage{amsmath,amssymb}
 
\begin{document}
\title{Theoretical investigation of a polarizable colloid in the salt medium }

\author{Amin Bakhshandeh}
\email{amin.bakhshandeh@ufrgs.br}
\affiliation{Instituto de F\'isica, Universidade Federal do Rio Grande do Sul, Caixa Postal 15051, CEP 91501-970, Porto Alegre, RS, Brazil}
 
\begin{abstract}
In the present work, we have extended a weak coupling theory [A. Bakhshandeh, A. P. dos Santos and Y. Levin Phys. Rev. Lett 107, 107801 (2011)] for systems with added 1:1 electrolyte. To study the accuracy of the developed theory, we compare its numerical predictions with Monte Carlo simulation data and a recent theory which accounts for the surface polarization. A very good agreement is found for the case of monovalent electrolytes, up to very high salt concentrations and different colloidal charges.

%\begin{keyword}
%colloid, Poisson-Boltzmann equation, ionic system 
%\end{keyword}

\end{abstract}

\maketitle
\section{Introduction}
Ionic systems always have been a central subject of Chemistry and Physics as well as Biology. There are very interesting and counterintuitive observations in ionic systems, such as attraction of like-charged macromolecules in solutions where multivalent counterions exist~\cite{adar,adar2016,levin,naji2004, Kanduc}. This observation has been detected in bacteriophage heads and some viruses  too~\cite{levin,bloo1,bloo2,klim,tang}. 
	
	Recently, Huang and Lapitsky have shown that monovalent salt increases colloidal stability when   Chitosan-Tripolyphosphate Microgels are forming, and they related this phenomenon with weakening chitosan-Tripolyphosphate binding ~\cite{Huang}.
	 It is also  well known that in biological systems maintenance of concentration of monovalent salt ions such as K and Na is very important for the cell stability, as the increase or decrease of the concentration of salt can lead to death of a cell ~\cite{ARINO20141}. 

Because of long-range interaction nature of these systems,the theoretical modeling of ionic systems is a rather challenging task. One of the most successful approaches to investigate these systems was proposed by Debye and H{\"u}ckel back in 1923~\cite{debye}. This theory allows us to calculate the chemical potential of positive and negative ions in the limit of very weak electrostatic couplings.
 
However, in the case of colloidal particles, the situation becomes more complicated. Since colloids have acidic
or basic charged groups on their surfaces, in a polar medium, these groups become ionized and the particles become charged~\cite{diehl2006}  and the DH theory looses its validity for such highly asymmetric systems. Repulsion between
like-charged colloidal particles keep them separated, however, it is well known that by addition of electrolyte to colloidal
suspension, at a critical concentration, the repulsive energy becomes weak enough, and coagulation takes place~\cite{Israel,levin,Cassou}. 

One successful theory which explains this phenomenon belongs to Derjaguin, Landau, Verwey, and Overbeek (DLVO) which accounts for the interaction between weakly charged homogeneous surface. It simply assumes that the net interaction between two colloidal surfaces is a combination of electrostatic double layer forces and van der Waals force~\cite{levin,verwey,derj,bakh2015,bakh2018}. 

The van der Waals force dominates
when the separation distance between the surfaces is small, while the electrostatic force becomes relevant on larger distances. The mentioned theory works reasonably well for weakly charged homogeneously surfaces and gives us a clear insight into coagulation phenomenon ~\cite{Israel,levin,bak2011,dos2011}.

Due to the strong variation of electric potential near a surface, electrical double layer (EDL) can be formed. EDL has a significant effect on the behavior of colloids in contact with solutions.
	Guerrero-Garc\'{\i}a \emph{et~al.}~\cite{g1,g2,g3} have shown the importance of coions in the electric double layer so that they are able to induce correlations which modify the EDL’s potential and this, for instance, causes an increase of the charge reversal phenomenon for 1: $\alpha$ salts ({\it{viz}}., univalent counterions and multivalent coions).

In general, in the presence of monovalent ions, charge renormalization stabilizes colloidal suspensions and phase separation does not happen in water~\cite{levin,marcia}. As we mentioned earlier, there are many chemical and biochemical systems which monovalent salt ions play important roles to stabilize them. Therefore, by studying the monovalent salt behavior around charged colloidal particles, we are able to evaluate the stability of colloidal suspensions in the mentioned systems that can be great of interest.

 In aqueous colloidal systems with monovalent counterions, the Poisson-Boltzmann (PB) equation is in very good agreement with the experiments and simulation data, but the PB
 equation does not account for the correlations between the counterions and breaks down for low dielectric solvents, when such correlations become strong~\cite{levin,naji,Trizac}. 
 
 Besides, 
 in many colloidal suspensions the dielectric constant of colloidal particles is smaller than that dielectric constant of the surrounding medium and, as a result, ions near colloidal surface feel a strong ion-image repulsion~\cite{dos2011}.
 The interaction between ions and image charges creates an extra force which can affect the effective charge of the colloidal particles. 
  This extra interaction leads to the modification of colloid-colloid and colloid-counterion interaction potentials~\cite{Israel,levin,messina,dos2011} and for this reason, for describing real systems, one should consider this issue. 
 
 However, since it is difficult to include the dielectric discontinuities in spherical or other complicated geometries, most of the theoretical and numerical works on colloidal suspensions ignore the effects of the dielectric discontinuity at the interface.

 One way to include polarizations in colloidal systems is by using the image-charge method~\cite{dos2011,Jackson}. Imagine a charge is put near a conducting sphere.
  To satisfy the Maxwell equations, one can assume that a charge is induced inside a sphere where its magnitude and position are dependent on the distance and charge magnitude of ion in medium~\cite{Jackson}.

By doing this, one can obtain the interaction between induced charges and counterions and carry out molecular dynamic or Monte Carlo (MC) simulations~\cite{dos2011}.
At this point, the question is, which kind of mean-field equation can  describe the distribution of counterions around a polarizable colloid?
Although the simple PB equation works very well for non-polarizable colloids in mean-field regions, it is not able to predict the effect of induced charges at the colloid-surface solvent interface.

A recent theory allows us to predict the counterion density distribution
in both weak and strong coupling limits~\cite{bak2011}. This theory is accurate for systems containing just counterions and without added salt.
In the present paper, we aim to extend this theory in the weak coupling region for a situation in which $1:1$ electrolyte is included.

In order to test the range of validity of the proposed approach, we test its predictions against MC simulation data for systems bearing different salt concentrations and colloidal surface charges. Moreover, we compare our extended theory with one recent theory which has been presented recently for surface polarizations problems~\cite{dos20181}.

This paper is organized as follows. In section II, we present the model and theoretical background, and show how we extend the theory for a case in which salt ions are present. 
In section III, we present and discuss simulation details and in section IV we summarize our results. Finally, in section VI
we drown the conclusions of our work.

\section{Theoretical background}
\begin{figure}
  \begin{center}
  \includegraphics[width=7cm]{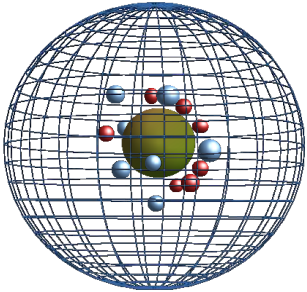}
  \end{center}
  \caption{ Wigner-Seitz cell. Colloid has been placed in center of the cell and ions are around it. }  
  \label{fig1}
  \end{figure}

The main part of the weak coupling theory relies on the mean-field PB equation. 
In the case of non-polarizable colloidal particles we know that
for monovalent counterions, the PB theory works very well~\cite{levin}. For polarizable
colloids, the usual PB equation fails even for
monovalent ions. 
This failure is due to the counterion-image interaction which is not considered in the mean-field PB equation. 
To satisfy Maxwell equations one can place an image charge inside the colloid at the inversion point 
and a counterimage line-charge distributed along the line connecting the center of the colloid and the image-charge~\cite{bak2011,dos2011,Jackson}.
Consider a colloidal particle with charge $Z $ and radius $a$ in  Wigner-Seitz (WS) cell of radius $R$, as is shown in Fig.~\ref{fig1}.
The starting point is the Poisson equation for an electrolyte $1:\alpha$ we have~\cite{bak2011,dos2011,adar2017bjerrum}:
%%%%%%%%%%%%%%%% equation %%%%%%%%%%%%%%%%%%%%%
\begin{equation}  
\nabla^2 \phi(r) = -\big(
\frac{Z q}{\epsilon_w a^2} \delta(r-a) + \frac{4 \pi q }{\epsilon_w } (\rho_+(r) -\alpha \rho_-(r))\big) ,
\label{Eq1}
\end{equation}   
%%%%%%%%%%%%% end of equation %%%%%%%%%%%%%%%%%
where $\phi(r)$, $a$, $\epsilon_w$, $q$, $\rho_\pm(r)$ and $\alpha$ are electrostatic potential, the radius of colloid, dielectric constant of water, the proton charge, density of positive and negative ions and ionic valence respectively. The ionic distributions,
$\rho(r)$, in the presence of $1:1$ salt ions, are defined as follows: 
%%%%%%%%%%%%%%%% equation %%%%%%%%%%%%%%%%%%%%%
\begin{equation}
\begin{split}
 \rho_-(r) = 
\frac{(Z+N_{s-})  e^{\beta (q \phi(r) -w(\kappa,z)) }}{4 \pi \int_{a+r_{ion}}^R dr r^2e^{\beta (q \phi(r) -w(\kappa,z) )} },\\
\\
 \rho_+(r) = 
\frac{N_{s+}~  e^{-\beta (q \phi(r) -w(\kappa,z)) }}{4 \pi \int_{a+r_{ion}}^R dr r^2e^{-\beta (q \phi(r) -w(\kappa,z) )} },
\label{Eq2}
\end{split}
\end{equation}
%%%%%%%%%%%%% end of equation %%%%%%%%%%%%%%%%%
where $N_{s+}$ and  $N_{s-}$ are the number of positive and negative ions, respectively, inside the cell, and $r_{ion}$ stands for the ionic radius. 
As can be seen in Eqs.~\ref{Eq2} there is a correction term $w(\kappa,z)$,  which is due to the effect of image charges.
As a matter of fact, in order to bring an ion close to the polarized surface one must perform two distinct works, one is against the surface charge of colloid and the other is against the induced image charges. This effect should be dependent on the Debye length and this correction can be obtained through the following relation:~\cite{levin21,bak2011}:
%%%%%%%%%%%%%%%% equation %%%%%%%%%%%%%%%%%%%%%
\begin{equation}
\begin{split}
w(\kappa,z) =
\big(k_B T \Delta(\kappa,z) +\\
\frac{r_{ion}}{z} \big(w_0(\kappa,r_{ion})-w_0(0,r_{ion})\big) 
\big)e^{-2 \kappa (z-r_{ion}) } ,
\label{Eq3}
\end{split}
\end{equation}
%%%%%%%%%%%%% end of equation %%%%%%%%%%%%%%%%%
where $k_B$, $T$ and $z$ are the Boltzmann constant, the
temperature and the distance from the colloidal surface respectively and $\kappa = \sqrt{8 \pi \lambda_B I}$, where $I$ is  ionic strength,  $\lambda_B=q^2/\epsilon_w k_B T$ is the Bjerrum length and  $w_0(\kappa,r_{ion})$ is given by
%%%%%%%%%%%%%%%% equation %%%%%%%%%%%%%%%%%%%%%
\begin{equation}
w_0(\kappa,r_{ion}) =k_B T \frac{\lambda_B \alpha^2}{2} \int_0^\infty dk\frac {k[s \cosh(k~r_{ion})-k \sinh(k~r_{ion})]}{s[s \cosh(k~r_{ion})+k \sinh(k~r_{ion})]} ,
\label{Eq4}
\end{equation}
%%%%%%%%%%%%% end of equation %%%%%%%%%%%%%%%%%
where $s = \sqrt{k^2+\kappa^2}$.
The first term in Eq.~\ref{Eq3} is due to spherical correction and the second term is due to broken transnational symmetry. Furthermore, it can be shown that the correction the term $\Delta(\kappa,z)$ is~\cite{bak2011} :
%%%%%%%%%%%%%%%%%%%%%%
\begin{equation}
\Delta(\kappa,z) = \frac{\alpha^2 a \lambda_B}{2(z^2+2az)} +\frac{\alpha^2 \lambda_B}{2 a}\log[1-\frac{a^2}{(z+a)^2)}] ,
\label{Eq43}
\end{equation}
%%%%%%%%%%%%% end of equation %%%%%%%%%%%%%%%%%
where the first term is due to the interaction between an ion and its image charge and the second term is due to ion-counterimage interaction.

As can be seen, when $a\rightarrow\infty$ the first term becomes a constant and the second term goes to zero. As a result, we expect that for large colloidal particles the correction term becomes less important, while for small size of colloidal particles the correction term becomes considerable.

It should be noted that Eqs.~\ref{Eq3}-\ref{Eq43} are obtained considering $\epsilon_c \approx 0 $. However, since for most of colloidal particles $\epsilon_w/\epsilon_c\gg1$, this approximation is well justified and we can therefore apply it to study theses systems.

  Now by solving Eqs.~\ref{Eq2} and ~\ref{Eq1} numerically in presence of salt ions the ionic density profiles around a polarizable colloid can be obtained.
 %%%%%%%%%%%%%%%% figure %%%%%%%%%%%%%%%%%%%%%
%%%%%%%%%%%%% end of figure %%%%%%%%%%%%%%%%%
  
\section{simulation details}
Although molecular dynamic and MC simulation of systems with dielectric mismatches are not straightforward, there are some recent simulations which have been applied to the planar and spherical geometry containing dielectric discontinuity ~\cite{monica,Wang,wang1,Wang2,gl}.
In the present study, we have used a method introduced in Ref.~\onlinecite{dos2011}.
	
\begin{figure}
  \begin{center}
  \includegraphics[width=7cm]{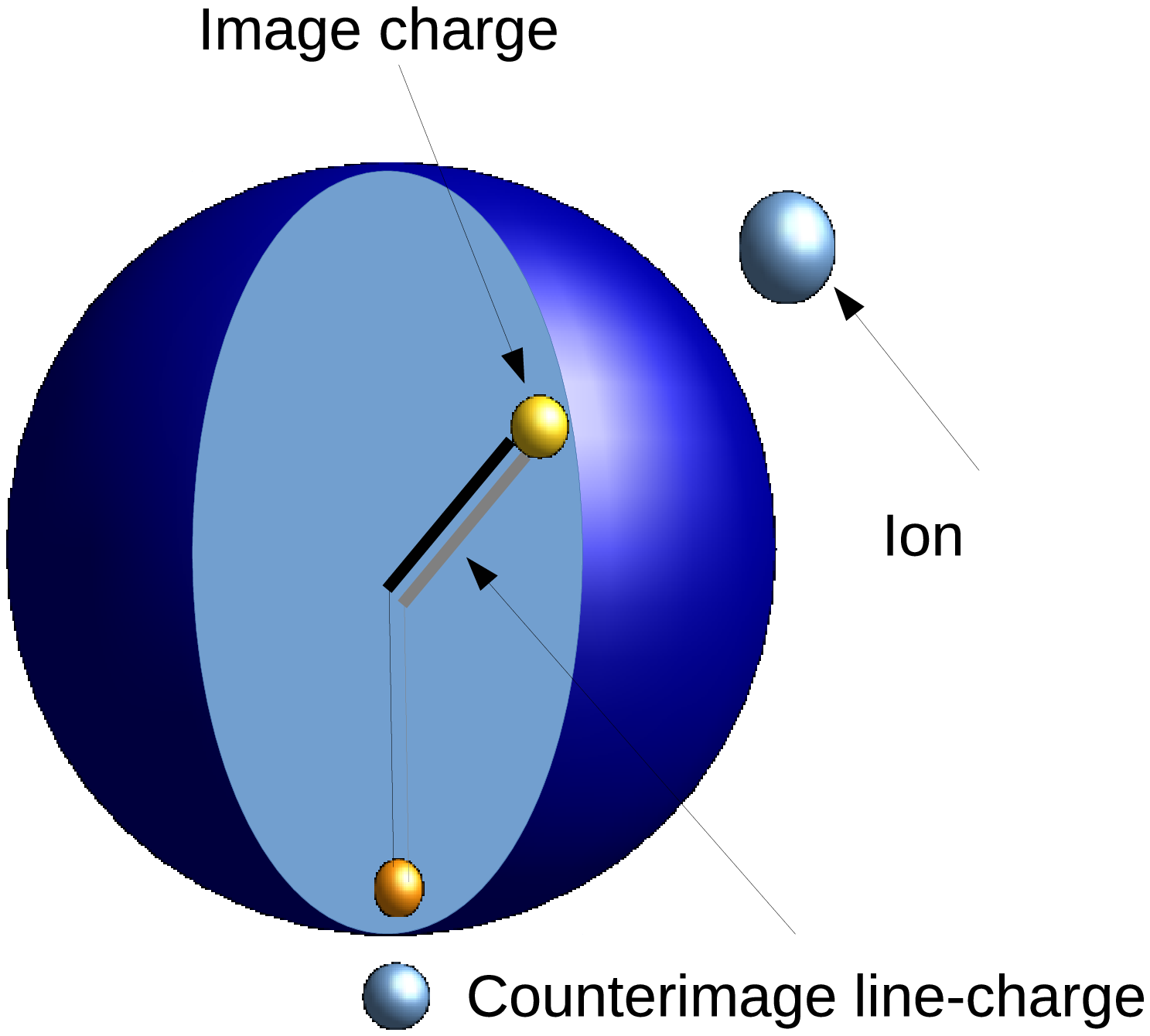}
  \end{center}
  \caption{Representation of ions near a colloidal particle and the induced images and counterimage charges. }  
  \label{fig1q}
  \end{figure}

	 From electrostatics we know that the tangential component of the electric field must be continuous through the colloid-media interface. 
	  It can be shown, that this boundary condition is satisfied by considering an image charge  which is located inside the colloid, as well as a counterimage line-charge, which is distributed along the line from the center of colloid to the position of image charge~\cite{Lindell,Norris} as shown in Fig.~\ref{fig1q}.
	The location $\bm{r}_{image}$ and charge magnitude $Q_{image}$ of image charge are dependent on position and charge of counterion as follows:
%%%%%%%%%%%%%	
	\begin{equation}
	\begin{split}
Q_{image} = \frac{\gamma\ Q a }{r_{ion}}	\\
\bm{r}_{image} = \frac{\bm{r}_{ion}~ a^2}{r_{ion}^2},
	 		\end{split}
	 		\label{Eq5}
	\end{equation}
%%%%%%%%%%%%%%
where $r_i$ and $Q$ are the position and charge of ions in media and   $\gamma = (\epsilon_w - \epsilon_c)/(\epsilon_w + \epsilon_c) $ is the relative dielectric mismatch, $\epsilon_c$ is dielectric constant of the colloidal particle.
Also, the line density of counterimage charge is not constant and is a function of  position~\cite{Norris,dos2011}. By using this information one can obtain the exact counterimage-ion interaction potential which is a hypergeometric function~\cite{dos2011}. However, by assuming that $\gamma \approx 1$, counterimage-ion potential at an arbitrary point simplifies to ~\cite{dos2011}:

	\begin{equation}
\begin{split}
\psi(\bm{r},\bm{r_i}) = \frac{\alpha\ q  }{\epsilon_w a  } \log{(\frac{r r_i - \bm{r} . \bm{r}_i}{a^2 -\bm{r}_i .\bm{r}+\sqrt{a^4 - 2 a^2 \bm{r} . \bm{r}_i +r^2r_i^2}})},
\end{split}
\label{Eq6}
\end{equation}
and also counterimage-ion interaction potential is:

	\begin{equation}
\begin{split}
\psi^{self}(\bm{r_i}) = \frac{\alpha\ q  }{\epsilon_w a  } \log{(1-\frac{a^2}{r_i^2})}.
\end{split}
\label{Eq7}
\end{equation}

The simulations in this study are performed inside a spherical Wigner-Seitz (WS) cell of radius $R$ with
a colloidal particle of charge $Zq$ placed at the center of the WS cell, as is depicted in Fig.~\ref{fig1}. In order to fulfill overall  electroneutrality, the cell also contains $ N = Z$ 
counterions. 

 We have placed 1:1 salt with density $\rho_0$ inside the cell. All ions have diameter 4 \AA. 
 By considering Eqs.~\ref{Eq6},\ref{Eq7} the total energy of the system can be calculated as follows:
 \begin{equation}
 \begin{split}
 U = \sum_{i=1}^{N} -\frac{Z \alpha q ^2}{\epsilon_w r_i} +\sum_{i=1}^{N} \bigg(\frac{\gamma \alpha^2 q^2 a}{2 \epsilon_w (r_i^2-a^2)} + \frac{\alpha q \gamma \psi^{self} (\bm{r}_i) }{2}\bigg) + \\  \sum_{i=1}^{N-1}\sum_{j=i+1}^{N} \alpha q \bigg(\frac{Z \alpha q ^2}{\epsilon_w |\bm{r}_j-\bm{r}_i|}+\frac{\gamma \alpha q  a }{\epsilon_w r_i|\bm{r}_j-\frac{a^2}{r^2_i}\bm{r}_i|} +\gamma \psi(\bm{r}_j,\bm{r}_i)\bigg).
 \end{split}
 \label{Eq8}
 \end{equation}

The solvent is assumed to be uniform, with dielectric of permittivity $\epsilon_w = 80$, for colloidal particle $\epsilon_c = 2$, and in this research we consider $\lambda_B = 7.2 $~\AA which corresponds to an aqueous solution at room temperature.

 To perform simulations, we use Eq.~\ref{Eq8} and a canonical Monte Carlo algorithm~\cite{Frenkel,Allen} with $10^8$ MC steps for equilibration and $10^5$ steps for production.

  We should note that, in general, ions in neighboring cells also create image charges inside the colloid, but considering the effect of other cells is a very difficult task. Nevertheless, in this research, we  ignore the effect of neighbor cells and as a result we do not need to use Ewald Summation techniques. 
 The details of the simulation method for polarizable colloids can be found in Ref.~\onlinecite{dos2011}.

 %%%%%%%%%%%%%%%% figure %%%%%%%%%%%%%%%%%%%%%
%%%%%%%%%%%%% end of figure %%%%%%%%%%%%%%%%%
%%%%%%%%%%%%%%%% equation %%%%%%%%%%%%%%%%%%%%%

%%%%%%%%%%%%% end of equation %%%%%%%%%%%%%%%%%
 \section{RESULTS and discussion}
There is a recent new theory describing polarizable colloids surfaces~\cite{dos20181}.
 In this new theory there is no spherical correction, because it is very difficult to adopt the correction to chaotropic ions. Since the goal of Ref.~\onlinecite{dos20181} was to keep the approximated potentials for chaotropes and kosmotropes ions at the same level, the correction term was ignored. In this paper, we compare our extended theory (ET) to Ref.~\onlinecite{dos20181} for  kosmotropes ions in order to check to what extend this correction affects the results.
 
\begin{figure}
  \begin{center}
  \includegraphics[width=7cm]{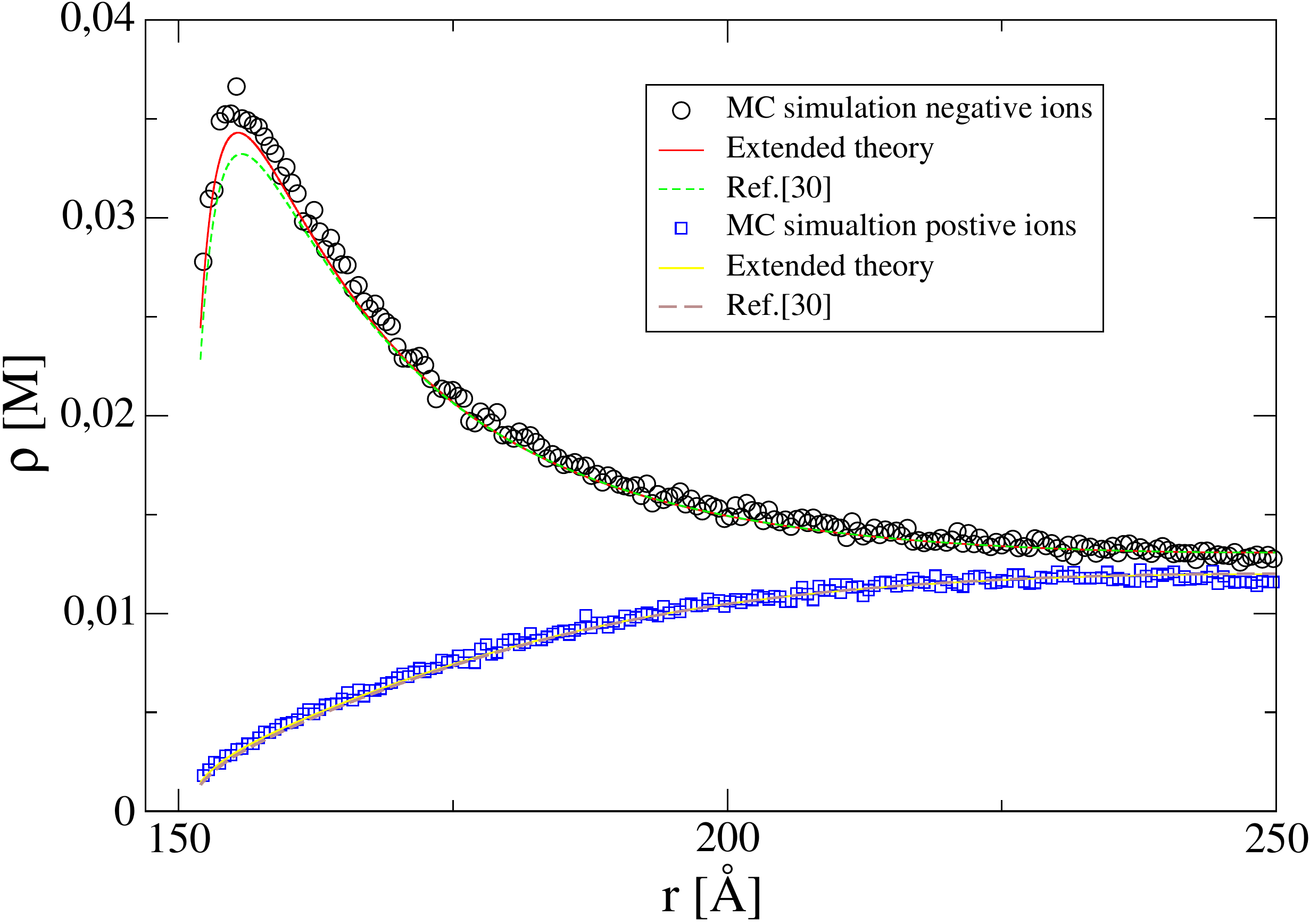}
  \end{center}
  \caption{Density profile for positive and negative ions, the radius of colloid is $150$ \AA~ and the salt concentration is $10$ mM, charge of colloid is $600~q$.  }  
  \label{fig2}
  \end{figure}

We start by studying a colloid with radius $150$ \AA, charge of $600 q$ and  salt concentration $10 $ mM. As can be seen in Fig~\ref{fig2} there is a very good agreement between the simulation data and (ET) results which is obtained by solving Eqs.~\ref{Eq2}. For negative ions there is a deviation between our ET and Ref.~\onlinecite{dos20181}, however, this deviation becomes small for positive ions. It seems that the correction term is less important for particle which are repelled from the surface.

 %%%%%%%%%%%%%%%% figure %%%%%%%%%%%%%%%%%%%%%
%%%%%%%%%%%%% end of figure %%%%%%%%%%%%%%%%%

\begin{figure}
  \begin{center}
  \includegraphics[width=7cm]{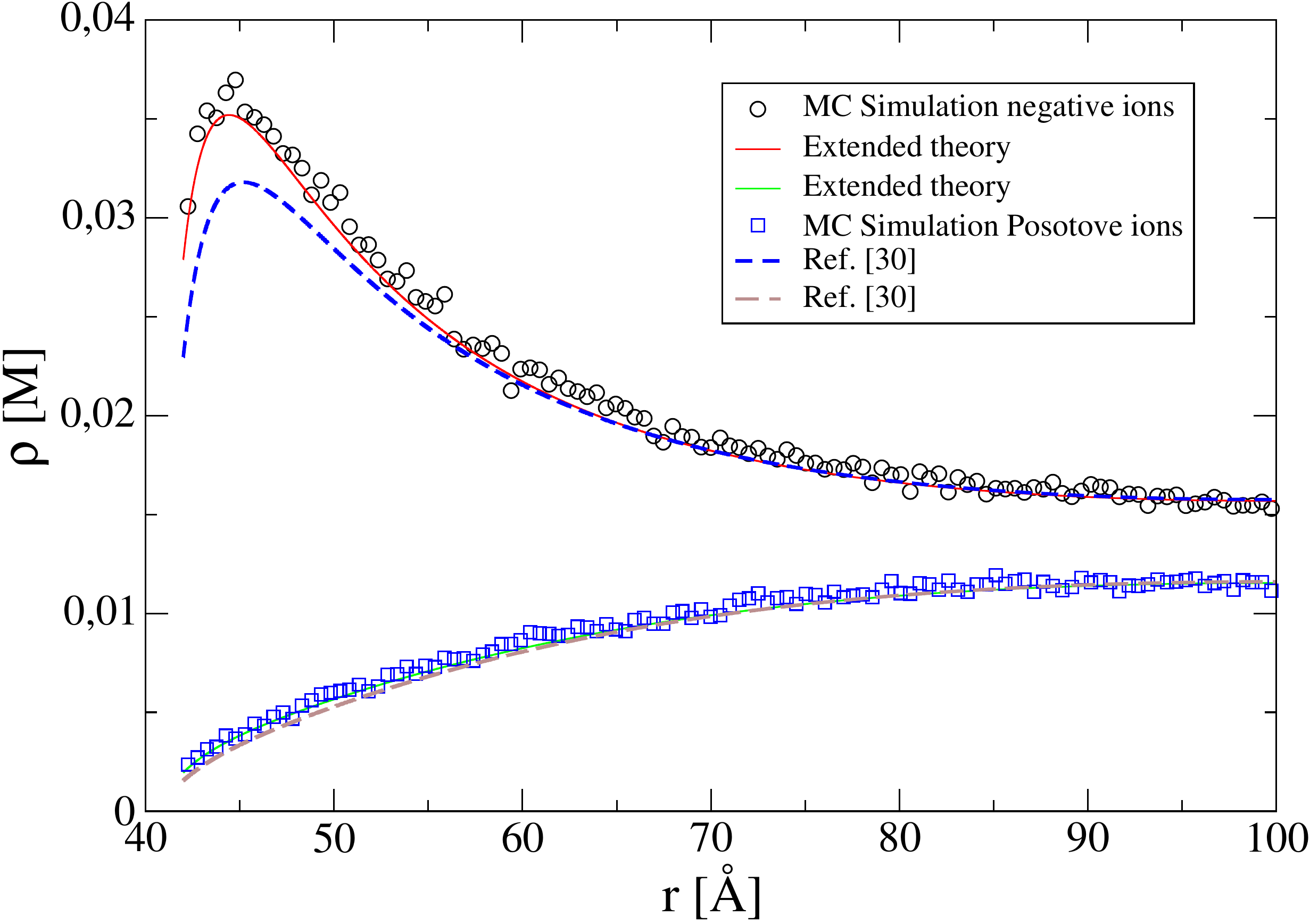}
  \end{center}
  \caption{Density profile for positive and negative ions. The radius of colloid is $40$ \AA~ and the salt concentration is $10$ mM charge of colloid is $20~q$.  }  
  \label{fig3}
  \end{figure}

We now reduce the radius of colloid to $40$ \AA, and its charge to $20~q$, and study the systems in presence of $10$ mM salt. As is seen in Fig.~\ref{fig3} the deviation between Ref.~\onlinecite{dos20181} and ET increases. The reason for this deviation is that, by reducing the colloidal size, the correction term becomes more important, in other words, the ions are more sensitive to changes of colloidal curvature when its radius become smaller. In contrast, the simulation data has great agreement with ET for positive ions while for negative ions the deviation between Ref.~\onlinecite{dos20181} and ET and simulation data is small.

\begin{figure}
  \begin{center}
  \includegraphics[width=7cm]{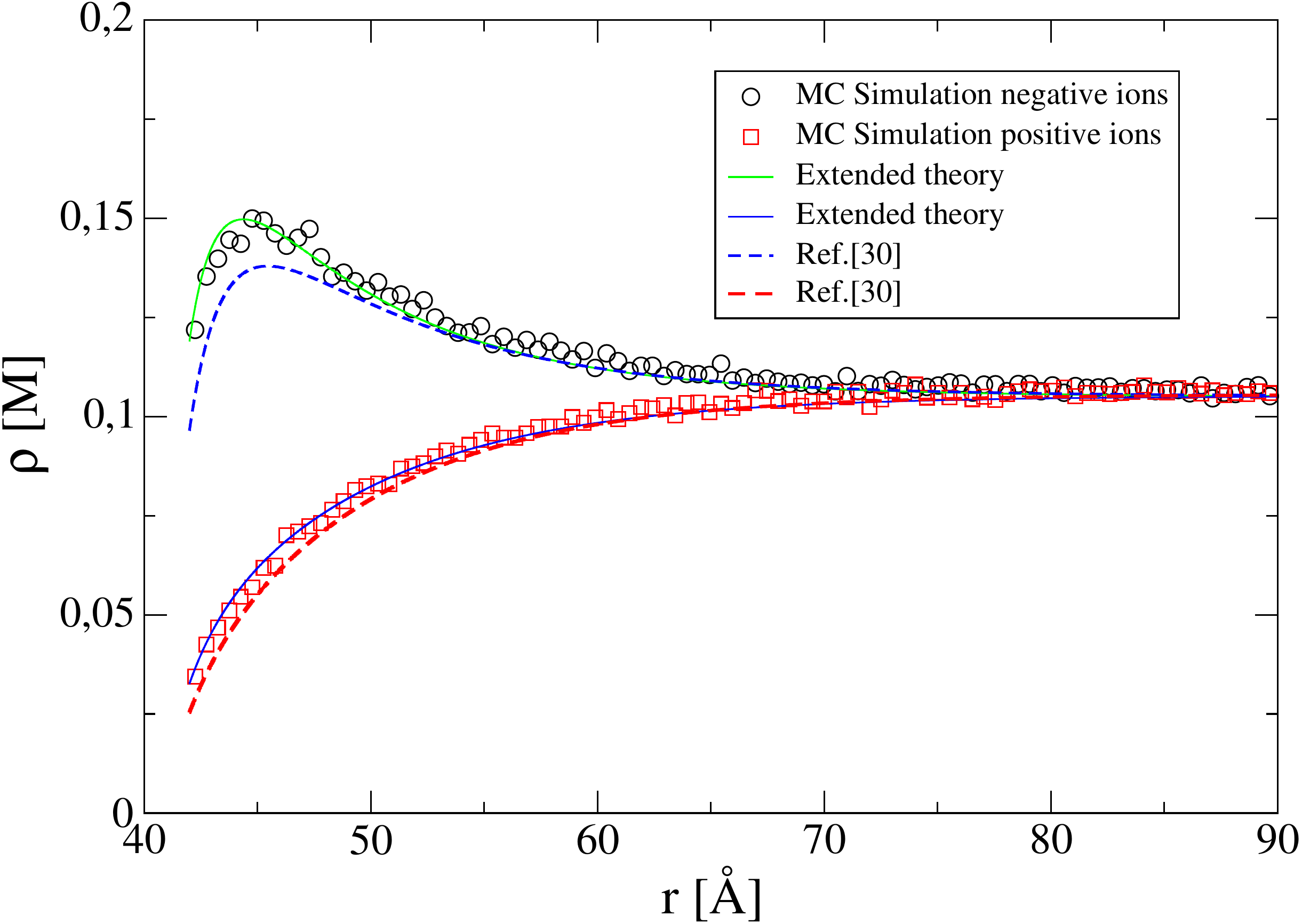}
  \end{center}
  \caption{Density profile for positive and negative ions. The radius of colloid is $40$ \AA~and the salt concentration is $100$ mM charge of colloid is $20~q$.  }  
  \label{fig4}
  \end{figure}

 By increasing the salt concentration to $100$ mM, the density profiles predicted by the two theories become more different. Again, as can be seen in Fig.~\ref{fig4}, the simulation data displays a very good agreement with the proposed ET. 
 
 As the concentration of salt increases, the number of negative particles near the colloidal surface increases too. In this case, the correction for ion-image interaction becomes more important and, as a result, one can see a more pronounced difference of predicted density profile for negative ions by these two theories. 
 Since ET uses the spherical correction it has better agreement with simulation data.

 %%%%%%%%%%%%%%%% figure %%%%%%%%%%%%%%%%%%%%%
%%%%%%%%%%%%% end of figure %%%%%%%%%%%%%%%%%

 %%%%%%%%%%%%%%%% figure %%%%%%%%%%%%%%%%%%%%%
\begin{figure}
  \begin{center}
  \includegraphics[width=7cm]{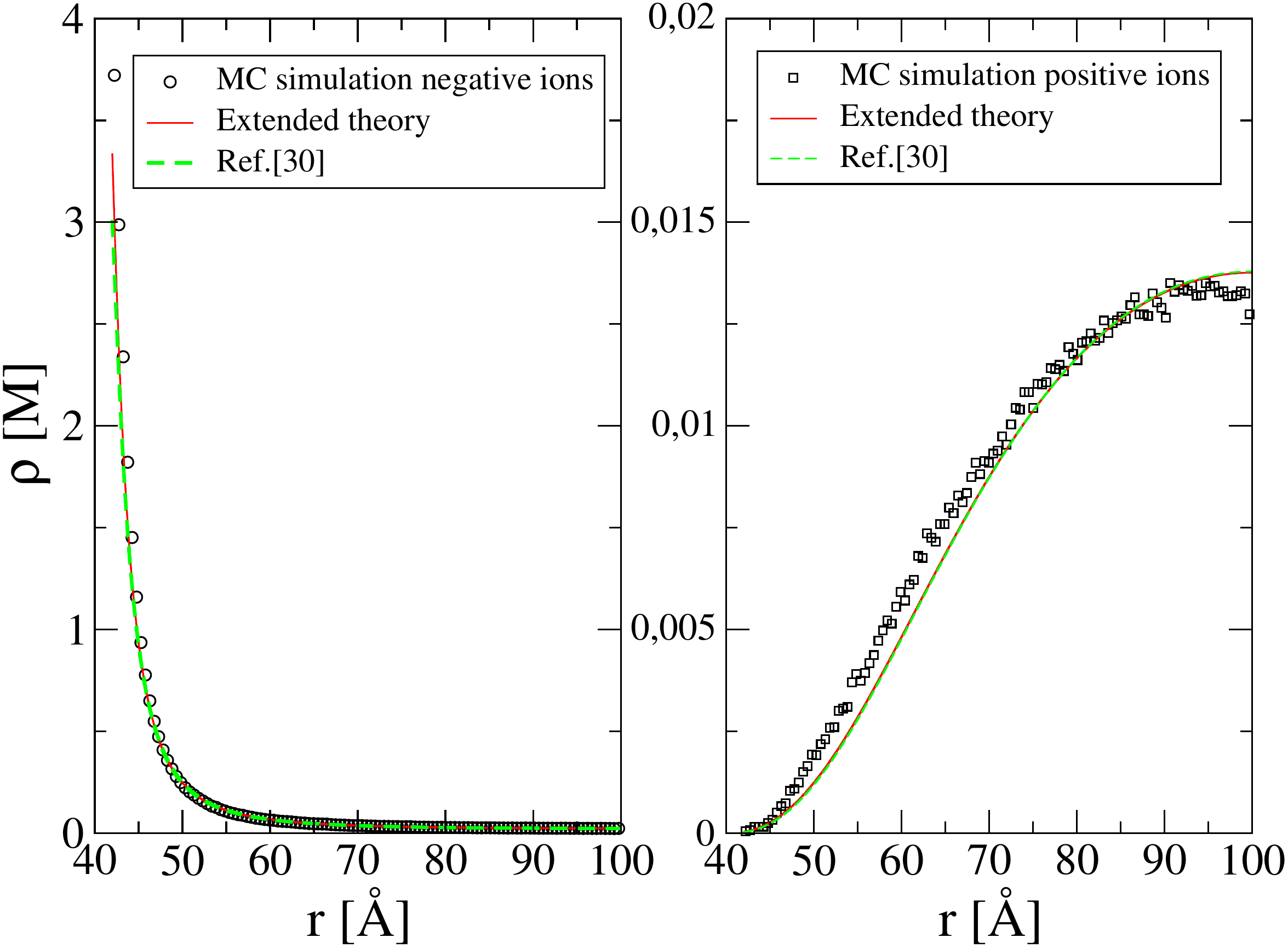}
  \end{center}
  \caption{Density profile for positive and negative ions. The radius of colloid is 40 \AA, the salt concentration is 10 mM and charge of colloid is 200.  }  
  \label{fig5}
  \end{figure}

%%%%%%%%%%%%% end of figure %%%%%%%%%%%%%%%%%
 We have also increased the charge of colloid to $200~q$, as can be seen in Fig.~\ref{fig5}. Now the maximum in the couterion density profile near the colloidal surface  vanishes. The reason for this observation is that, the image charge repulsion is not strong enough to repel ions from the surface because of the high  attraction of ions to colloidal surface. 
 %%%%%%%%%%%%%%%% figure %%%%%%%%%%%%%%%%%%%%%
%%%%%%%%%%%%% end of figure %%%%%%%%%%%%%%%%%
For this case, as can be seen, the agreement between the ET and Ref.~\onlinecite{dos20181} is very good even near the colloidal surface. However, the results from Ref.~\onlinecite{dos20181} predict much less adsorption of ions on the surface, but in the other hand, for positive ions the deviation between the two theories become very small.

There are many general theories for polarizable colloids in salt environments which are not restricted to
the weak coupling limit~\cite{monica,naji,Wang,wang1,Wang2,gl}. As a result, it is important to observe the performance of the ET for multivalence salt. Our extended mean-field theory, collapses for salt $1:3$ even for the small amount of salt concentration, for this reason we limit ourselves to salt $1:2$ 

However, we study the system in presence of salt $1:2$ for two salt concentrations of $10$ and $100$ mM. We consider a colloidal particle with radius $30$ \AA~and with a charge of $12~q$. 

\begin{figure}
 	\begin{center}
 		\includegraphics[width=7cm]{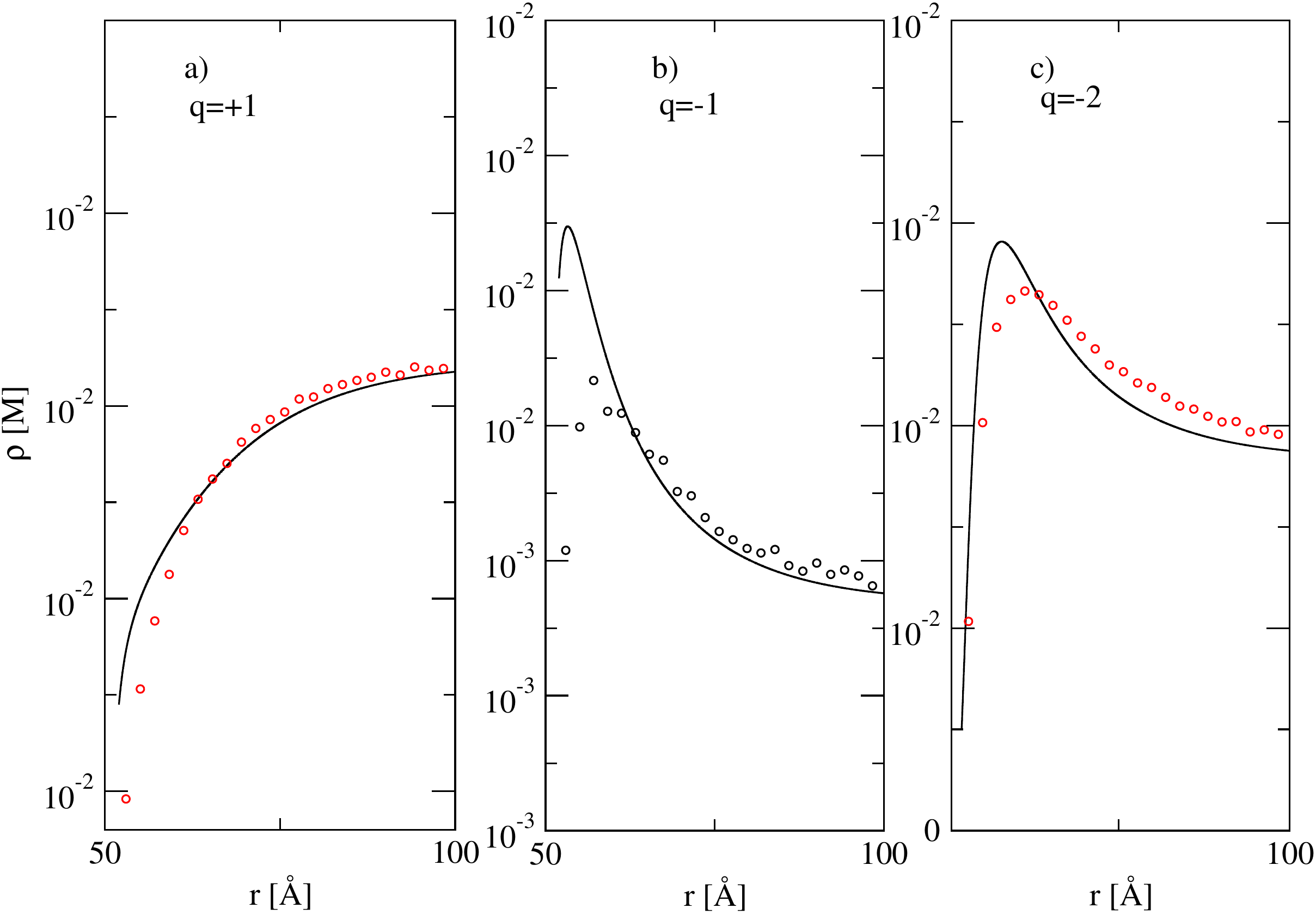}
 	\end{center}
 	\caption{Density profile for  counterions and ions of $1:2$ salt. The radius of colloid is $30$ \AA~with charge $12~q$ and the salt concentration is $10$ mM. The solid lines are predictions from the ET and circles are simulations data.  a) Positive ions of the salt. b) Negative dissociated counterions. c) Negative ions of salt.   }  
 	\label{fig6}
 \end{figure}

For the case of $10$ mM, as can be seen in Fig.~\ref{fig6}, the ET reasonably predicts the behavior of positive ions. For ions with charge $-2~q$ there is a deviation between ET and simulation data. For the monovalent counterions, however, the theory is able to predict reasonably well the ionic profiles far away from the surface, although the agreement breaks down at shorter distances.   

\begin{figure}
	\begin{center}
		\includegraphics[width=7cm]{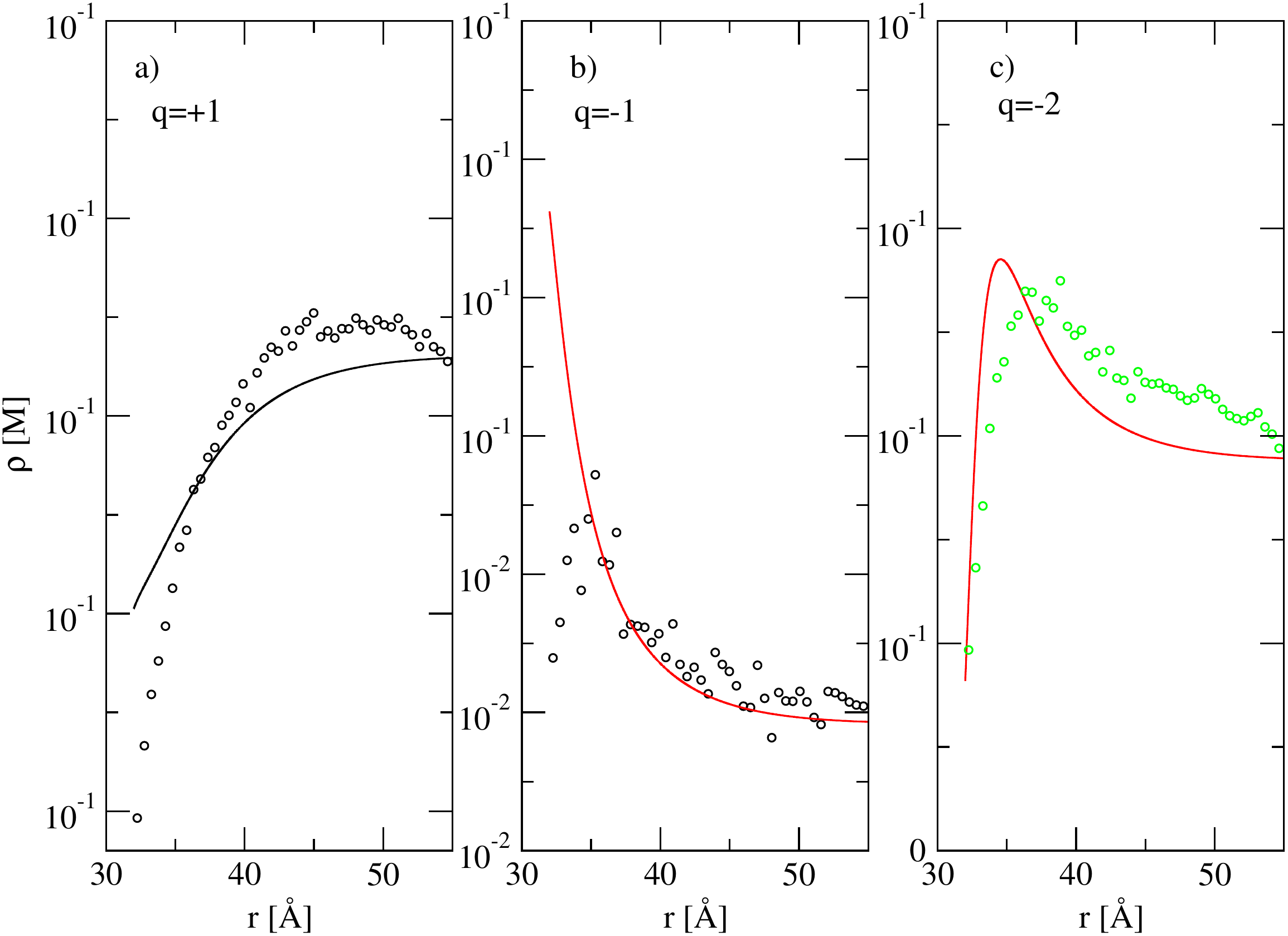}
	\end{center}
	\caption{Density profile for  dissociated counterions and the ions of $1:2$ salt. The radius of colloid is $30$ \AA~with charge $12~q$ and the salt concentration is $100$ mM. The solid lines are predictions from the ET and circles are simulations data.  a) Positive ions of the salt. b) Negative dissociated counterions. c) Negative ions of salt.  }  
	\label{fig7}
\end{figure}

By increasing the concentration of  $1:2$ salt to $100$ mM,  the theory breaks down as is seen in Fig.~\ref{fig7}. This result is not surprising since the proposed ET is a mean-field theory.  By increasing the ionic correlations the ET becomes therefore unable to predict the behavior of the system.

Finally, to check the performance of the ET at high concentration of monovalent salt and also to make systems similar to models that have been considered in some recent theoretical works such as Ref.~\onlinecite{gl}, we consider two cases. We keep the radius of colloid $30$ \AA~for two cases but we change the colloidal charge and concentration of $1:1$ salt. 

\begin{figure}
 	\begin{center}
 		\includegraphics[width=7cm]{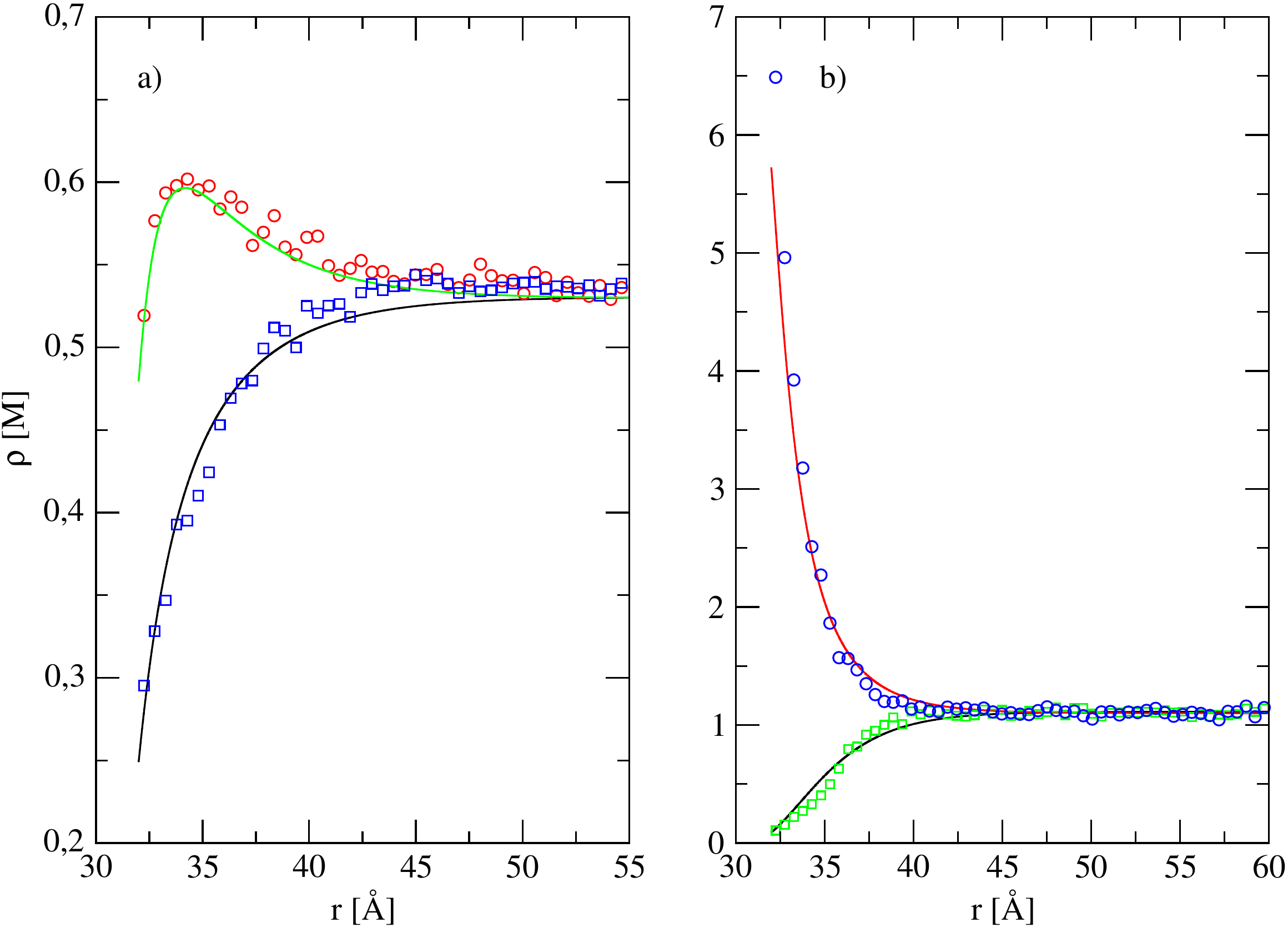}
 	\end{center}
 	\caption{Density profiles for ions of $1:1$ salt. The solid lines are predictions from the ET and circles are simulations data. a) Case I, the radius of colloid is $30$ \AA~ with charge $12~q$ and salt concentration  $500$ mM.   b) Case II, the radius of the colloid is $30$ \AA~with charge $120~q$ and salt concentration  $1$ M.   }  
 	\label{fig8}
 \end{figure}

In case I, the charge of colloid is $12~q$ and the monovalent salt concentration is $500$ mM. As is seen in Fig.~\ref{fig8} there is a perfect agreement between simulation data and ET. 

In the case of II, we increase the charge of colloid to $120~q$ and also salt concentration to $1$ M. As can be seen in  Fig.~\ref{fig8}  again it is observed that even for such high salt concentrations the ET works very well.
 
%%%%%%%%%%%%%%%%%%%%%%%%%%
%%%%%%%%%%%%%%%%%%%%%%%%%%
%%%%%%%%%%%%%%%%%%%%%%%%%%
 %%%%%%%%%%%%%%%%%%%%%%%%%%
 %%%%%%%%%%%%%%%%%%%%%%%%%%
 \section{Conclusions}
In the present work, we have developed a weak coupling theory for colloidal particles in the presence of added salt. 
To test the accuracy of the extended theory we compared our result with simulation data as well as with one recent theory which does not have the spherical correction.

 To this end, we studied a polarizable colloid with the positive charge and observed that the theory works very well for very high surface charge densities of colloidal particles. Also the ET's performance for biological monovalent salt concentration which is around $100$ mM is very good.
 
  We observed that the correction term normally has less effect on positive ions, but in the case of negative ions this correction becomes more relevant, especially for smaller colloidal radius where curvature effects become more pronounced. 
 	
 	 For high charged colloidal particle, the predicted results for positive and negative ions become very close, but for negative ions, Ref.~\onlinecite{dos20181} predicts less adsorption of ions on the colloidal surface.
 	
In this investigation, we further tested the predictions of ET for multivalent salt ions and we observed that even for low $2:1$ salt concentration, the ET has poor performance. The reason is that the basis of ET is PB equation which works well only in mean-field regions, and by increasing the strength of ionic correlations the mean-field picture looses its validity.
 
 It was observed that the ET works very well for a high concentration of $1:1$ salt ($500$ mM and 1 M) with a high surface charge density of colloidal particle.

In the end, by comparison with the theory presented in Ref.~\onlinecite{dos20181} it is seen that, the correction term $\Delta(\kappa,z)$  which is a combination of interaction of ion-image charge and ion-counterimage (which is an uniform linearly distributed charge) plays a very important role for small size of colloidal particles, On the other hand, by increasing the radius of the colloidal particle this contribution goes to a constant value which can be ignored.

 In our future work, we aim to also extend the strong coupling theory to salt medium.

%%%%%%%%%%%%% end of figure %%%%%%%%%%%%%%%%%
\section{ACKNOWLEDGMENT}
This work was supported by the Brazilian agency CNPq.
A.B thanks A. P. dos Santos for great discussion during preparation of this paper. Also A. B indebted to Martin Barmatz and Thiago E Colla for critical reading  of the manuscript.
 %%%%%%%%%%%%%%%% figure %%%%%%%%%%%%%%%%%%%%%
 %\begin{figure}
 %\begin{center}
 %\includegraphics[width=7cm]{zgraph1.eps}
 %\end{center}
 %\caption{Density profile of positive and negative ions near the charged plate for the case AII. }  
 %\label{fig5}
 %\end{figure}
%%%%%%%%%%%%% end of figure %%%%%%%%%%%%%%%%%
  %%%%%%%%%%%%%%%% figure %%%%%%%%%%%%%%%%%%%%%
 
%%%%%%%%%%%%% end of figure %%%%%%%%%%%%%%%%%
 
\bibliography{ref}

%merlin.mbs aipnum4-1.bst 2010-07-25 4.21a (PWD, AO, DPC) hacked
%Control: key (0)
%Control: author (8) initials jnrlst
%Control: editor formatted (1) identically to author
%Control: production of article title (-1) disabled
%Control: page (0) single
%Control: year (1) truncated
%Control: production of eprint (0) enabled
\begin{thebibliography}{41}%
\makeatletter
\providecommand \@ifxundefined [1]{%
 \@ifx{#1\undefined}
}%
\providecommand \@ifnum [1]{%
 \ifnum #1\expandafter \@firstoftwo
 \else \expandafter \@secondoftwo
 \fi
}%
\providecommand \@ifx [1]{%
 \ifx #1\expandafter \@firstoftwo
 \else \expandafter \@secondoftwo
 \fi
}%
\providecommand \natexlab [1]{#1}%
\providecommand \enquote  [1]{``#1''}%
\providecommand \bibnamefont  [1]{#1}%
\providecommand \bibfnamefont [1]{#1}%
\providecommand \citenamefont [1]{#1}%
\providecommand \href@noop [0]{\@secondoftwo}%
\providecommand \href [0]{\begingroup \@sanitize@url \@href}%
\providecommand \@href[1]{\@@startlink{#1}\@@href}%
\providecommand \@@href[1]{\endgroup#1\@@endlink}%
\providecommand \@sanitize@url [0]{\catcode `\\12\catcode `\$12\catcode
  `\&12\catcode `\#12\catcode `\^12\catcode `\_12\catcode `\%12\relax}%
\providecommand \@@startlink[1]{}%
\providecommand \@@endlink[0]{}%
\providecommand \url  [0]{\begingroup\@sanitize@url \@url }%
\providecommand \@url [1]{\endgroup\@href {#1}{\urlprefix }}%
\providecommand \urlprefix  [0]{URL }%
\providecommand \Eprint [0]{\href }%
\providecommand \doibase [0]{http://dx.doi.org/}%
\providecommand \selectlanguage [0]{\@gobble}%
\providecommand \bibinfo  [0]{\@secondoftwo}%
\providecommand \bibfield  [0]{\@secondoftwo}%
\providecommand \translation [1]{[#1]}%
\providecommand \BibitemOpen [0]{}%
\providecommand \bibitemStop [0]{}%
\providecommand \bibitemNoStop [0]{.\EOS\space}%
\providecommand \EOS [0]{\spacefactor3000\relax}%
\providecommand \BibitemShut  [1]{\csname bibitem#1\endcsname}%
\let\auto@bib@innerbib\@empty
%</preamble>
\bibitem [{\citenamefont {Adar}, \citenamefont {Andelman},\ and\ \citenamefont
  {Diamant}(2017)}]{adar}%
  \BibitemOpen
  \bibfield  {author} {\bibinfo {author} {\bibfnamefont {R.~M.}\ \bibnamefont
  {Adar}}, \bibinfo {author} {\bibfnamefont {D.}~\bibnamefont {Andelman}}, \
  and\ \bibinfo {author} {\bibfnamefont {H.}~\bibnamefont {Diamant}},\
  }\href@noop {} {\bibfield  {journal} {\bibinfo  {journal} {Advances in
  colloid and interface science}\ }\textbf {\bibinfo {volume} {247}},\ \bibinfo
  {pages} {198} (\bibinfo {year} {2017})}\BibitemShut {NoStop}%
\bibitem [{\citenamefont {Adar}, \citenamefont {Andelman},\ and\ \citenamefont
  {Diamant}(2016)}]{adar2016}%
  \BibitemOpen
  \bibfield  {author} {\bibinfo {author} {\bibfnamefont {R.~M.}\ \bibnamefont
  {Adar}}, \bibinfo {author} {\bibfnamefont {D.}~\bibnamefont {Andelman}}, \
  and\ \bibinfo {author} {\bibfnamefont {H.}~\bibnamefont {Diamant}},\
  }\href@noop {} {\bibfield  {journal} {\bibinfo  {journal} {Physical Review
  E}\ }\textbf {\bibinfo {volume} {94}},\ \bibinfo {pages} {022803} (\bibinfo
  {year} {2016})}\BibitemShut {NoStop}%
\bibitem [{\citenamefont {Levin}(2002)}]{levin}%
  \BibitemOpen
  \bibfield  {author} {\bibinfo {author} {\bibfnamefont {Y.}~\bibnamefont
  {Levin}},\ }\href@noop {} {\bibfield  {journal} {\bibinfo  {journal} {Reports
  on progress in physics}\ }\textbf {\bibinfo {volume} {65}},\ \bibinfo {pages}
  {1577} (\bibinfo {year} {2002})}\BibitemShut {NoStop}%
\bibitem [{\citenamefont {Naji}\ and\ \citenamefont {Netz}(2004)}]{naji2004}%
  \BibitemOpen
  \bibfield  {author} {\bibinfo {author} {\bibfnamefont {A.}~\bibnamefont
  {Naji}}\ and\ \bibinfo {author} {\bibfnamefont {R.~R.}\ \bibnamefont
  {Netz}},\ }\href@noop {} {\bibfield  {journal} {\bibinfo  {journal} {The
  European Physical Journal E}\ }\textbf {\bibinfo {volume} {13}},\ \bibinfo
  {pages} {43} (\bibinfo {year} {2004})}\BibitemShut {NoStop}%
\bibitem [{\citenamefont {Petersen}\ \emph {et~al.}(2018)\citenamefont
  {Petersen}, \citenamefont {Roa}, \citenamefont {Dzubiella},\ and\
  \citenamefont {Kanduc}}]{Kanduc}%
  \BibitemOpen
  \bibfield  {author} {\bibinfo {author} {\bibfnamefont {B.}~\bibnamefont
  {Petersen}}, \bibinfo {author} {\bibfnamefont {R.}~\bibnamefont {Roa}},
  \bibinfo {author} {\bibfnamefont {J.}~\bibnamefont {Dzubiella}}, \ and\
  \bibinfo {author} {\bibfnamefont {M.}~\bibnamefont {Kanduc}},\ }\href
  {\doibase 10.1039/C8SM00399H} {\bibfield  {journal} {\bibinfo  {journal}
  {Soft Matter}\ }\textbf {\bibinfo {volume} {14}},\ \bibinfo {pages} {4053}
  (\bibinfo {year} {2018})}\BibitemShut {NoStop}%
\bibitem [{\citenamefont {Bloomfield}(1991)}]{bloo1}%
  \BibitemOpen
  \bibfield  {author} {\bibinfo {author} {\bibfnamefont {V.~A.}\ \bibnamefont
  {Bloomfield}},\ }\href@noop {} {\bibfield  {journal} {\bibinfo  {journal}
  {Biopolymers}\ }\textbf {\bibinfo {volume} {31}},\ \bibinfo {pages} {1471}
  (\bibinfo {year} {1991})}\BibitemShut {NoStop}%
\bibitem [{\citenamefont {Bloomfield}(1997)}]{bloo2}%
  \BibitemOpen
  \bibfield  {author} {\bibinfo {author} {\bibfnamefont {V.~A.}\ \bibnamefont
  {Bloomfield}},\ }\href@noop {} {\bibfield  {journal} {\bibinfo  {journal}
  {Biopolymers}\ }\textbf {\bibinfo {volume} {44}},\ \bibinfo {pages} {269}
  (\bibinfo {year} {1997})}\BibitemShut {NoStop}%
\bibitem [{\citenamefont {Klimenko}, \citenamefont {Tikchonenko},\ and\
  \citenamefont {Andreev}(1967)}]{klim}%
  \BibitemOpen
  \bibfield  {author} {\bibinfo {author} {\bibfnamefont {S.}~\bibnamefont
  {Klimenko}}, \bibinfo {author} {\bibfnamefont {T.}~\bibnamefont
  {Tikchonenko}}, \ and\ \bibinfo {author} {\bibfnamefont {V.}~\bibnamefont
  {Andreev}},\ }\href@noop {} {\bibfield  {journal} {\bibinfo  {journal}
  {Journal of molecular biology}\ }\textbf {\bibinfo {volume} {23}},\ \bibinfo
  {pages} {523} (\bibinfo {year} {1967})}\BibitemShut {NoStop}%
\bibitem [{\citenamefont {Tang}\ \emph {et~al.}(1996)\citenamefont {Tang},
  \citenamefont {Wong}, \citenamefont {Tran},\ and\ \citenamefont
  {Janmey}}]{tang}%
  \BibitemOpen
  \bibfield  {author} {\bibinfo {author} {\bibfnamefont {J.~X.}\ \bibnamefont
  {Tang}}, \bibinfo {author} {\bibfnamefont {S.}~\bibnamefont {Wong}}, \bibinfo
  {author} {\bibfnamefont {P.~T.}\ \bibnamefont {Tran}}, \ and\ \bibinfo
  {author} {\bibfnamefont {P.~A.}\ \bibnamefont {Janmey}},\ }\href@noop {}
  {\bibfield  {journal} {\bibinfo  {journal} {Berichte der Bunsengesellschaft
  f{\"u}r physikalische Chemie}\ }\textbf {\bibinfo {volume} {100}},\ \bibinfo
  {pages} {796} (\bibinfo {year} {1996})}\BibitemShut {NoStop}%
\bibitem [{\citenamefont {Huang}\ and\ \citenamefont {Lapitsky}(2011)}]{Huang}%
  \BibitemOpen
  \bibfield  {author} {\bibinfo {author} {\bibfnamefont {Y.}~\bibnamefont
  {Huang}}\ and\ \bibinfo {author} {\bibfnamefont {Y.}~\bibnamefont
  {Lapitsky}},\ }\href@noop {} {\bibfield  {journal} {\bibinfo  {journal}
  {Langmuir}\ }\textbf {\bibinfo {volume} {27}},\ \bibinfo {pages} {10392}
  (\bibinfo {year} {2011})}\BibitemShut {NoStop}%
\bibitem [{\citenamefont {Ariño}\ \emph {et~al.}(2014)\citenamefont {Ariño},
  \citenamefont {Aydar}, \citenamefont {Drulhe}, \citenamefont {Ganser},
  \citenamefont {Jorrín}, \citenamefont {Kahm}, \citenamefont {Krause},
  \citenamefont {Petrezsélyová}, \citenamefont {Yenush}, \citenamefont
  {Zimmermannová}, \citenamefont {van Heusden}, \citenamefont {Kschischo},
  \citenamefont {Ludwig}, \citenamefont {Palmer}, \citenamefont {Ramos},\ and\
  \citenamefont {Sychrová}}]{ARINO20141}%
  \BibitemOpen
  \bibfield  {author} {\bibinfo {author} {\bibfnamefont {J.}~\bibnamefont
  {Ariño}}, \bibinfo {author} {\bibfnamefont {E.}~\bibnamefont {Aydar}},
  \bibinfo {author} {\bibfnamefont {S.}~\bibnamefont {Drulhe}}, \bibinfo
  {author} {\bibfnamefont {D.}~\bibnamefont {Ganser}}, \bibinfo {author}
  {\bibfnamefont {J.}~\bibnamefont {Jorrín}}, \bibinfo {author} {\bibfnamefont
  {M.}~\bibnamefont {Kahm}}, \bibinfo {author} {\bibfnamefont {F.}~\bibnamefont
  {Krause}}, \bibinfo {author} {\bibfnamefont {S.}~\bibnamefont
  {Petrezsélyová}}, \bibinfo {author} {\bibfnamefont {L.}~\bibnamefont
  {Yenush}}, \bibinfo {author} {\bibfnamefont {O.}~\bibnamefont
  {Zimmermannová}}, \bibinfo {author} {\bibfnamefont {G.~P.~H.}\ \bibnamefont
  {van Heusden}}, \bibinfo {author} {\bibfnamefont {M.}~\bibnamefont
  {Kschischo}}, \bibinfo {author} {\bibfnamefont {J.}~\bibnamefont {Ludwig}},
  \bibinfo {author} {\bibfnamefont {C.}~\bibnamefont {Palmer}}, \bibinfo
  {author} {\bibfnamefont {J.}~\bibnamefont {Ramos}}, \ and\ \bibinfo {author}
  {\bibfnamefont {H.}~\bibnamefont {Sychrová}},\ }in\ \href {\doibase
  https://doi.org/10.1016/B978-0-12-800143-1.00001-4} {\emph {\bibinfo
  {booktitle} {Advances in Microbial Systems Biology}}},\ \bibinfo {series}
  {Advances in Microbial Physiology}, Vol.~\bibinfo {volume} {64},\ \bibinfo
  {editor} {edited by\ \bibinfo {editor} {\bibfnamefont {R.~K.}\ \bibnamefont
  {Poole}}}\ (\bibinfo  {publisher} {Academic Press},\ \bibinfo {year} {2014})\
  pp.\ \bibinfo {pages} {1 -- 63}\BibitemShut {NoStop}%
\bibitem [{\citenamefont {Debye}\ and\ \citenamefont
  {H{\"u}ckel}(1923)}]{debye}%
  \BibitemOpen
  \bibfield  {author} {\bibinfo {author} {\bibfnamefont {P.}~\bibnamefont
  {Debye}}\ and\ \bibinfo {author} {\bibfnamefont {E.}~\bibnamefont
  {H{\"u}ckel}},\ }\href@noop {} {\bibfield  {journal} {\bibinfo  {journal}
  {Physikalische Zeitschrift}\ }\textbf {\bibinfo {volume} {24}},\ \bibinfo
  {pages} {185} (\bibinfo {year} {1923})}\BibitemShut {NoStop}%
\bibitem [{\citenamefont {Diehl}\ and\ \citenamefont
  {Levin}(2006)}]{diehl2006}%
  \BibitemOpen
  \bibfield  {author} {\bibinfo {author} {\bibfnamefont {A.}~\bibnamefont
  {Diehl}}\ and\ \bibinfo {author} {\bibfnamefont {Y.}~\bibnamefont {Levin}},\
  }\href@noop {} {\bibfield  {journal} {\bibinfo  {journal} {The Journal of
  chemical physics}\ }\textbf {\bibinfo {volume} {125}},\ \bibinfo {pages}
  {054902} (\bibinfo {year} {2006})}\BibitemShut {NoStop}%
\bibitem [{\citenamefont {{J Israelachvili}}(1991)}]{Israel}%
  \BibitemOpen
  \bibfield  {author} {\bibinfo {author} {\bibnamefont {{J Israelachvili}}},\
  }\href@noop {} {\emph {\bibinfo {title} {Intermolecular and Surface
  Forces}}}\ (\bibinfo  {publisher} {Academic Press},\ \bibinfo {address}
  {London},\ \bibinfo {year} {1991})\BibitemShut {NoStop}%
\bibitem [{\citenamefont {{M Lozada-Cassou and R Saavedra-Barrera and D Hen-
  derson}}(1982)}]{Cassou}%
  \BibitemOpen
  \bibfield  {author} {\bibinfo {author} {\bibnamefont {{M Lozada-Cassou and R
  Saavedra-Barrera and D Hen- derson}}},\ }\href@noop {} {\bibfield  {journal}
  {\bibinfo  {journal} {J. Chem. Phys}\ }\textbf {\bibinfo {volume} {77}},\
  \bibinfo {pages} {5150} (\bibinfo {year} {1982})}\BibitemShut {NoStop}%
\bibitem [{\citenamefont {Verwey}\ and\ \citenamefont
  {Overbeek}(1948)}]{verwey}%
  \BibitemOpen
  \bibfield  {author} {\bibinfo {author} {\bibfnamefont {J.~W.}\ \bibnamefont
  {Verwey}}\ and\ \bibinfo {author} {\bibfnamefont {J.~G.}\ \bibnamefont
  {Overbeek}},\ }\href@noop {} {\emph {\bibinfo {title} {Theory of Stability of
  Lyophobic Colloids' Elsevier Publishing" Co}}}\ (\bibinfo  {publisher}
  {Amsterdam Holland},\ \bibinfo {year} {1948})\BibitemShut {NoStop}%
\bibitem [{\citenamefont {Derjaguin}(1941)}]{derj}%
  \BibitemOpen
  \bibfield  {author} {\bibinfo {author} {\bibfnamefont {B.~V.}\ \bibnamefont
  {Derjaguin}},\ }\href@noop {} {\bibfield  {journal} {\bibinfo  {journal}
  {Acta Physicochim. URSS}\ }\textbf {\bibinfo {volume} {14}},\ \bibinfo
  {pages} {633} (\bibinfo {year} {1941})}\BibitemShut {NoStop}%
\bibitem [{\citenamefont {Bakhshandeh}\ \emph {et~al.}(2015)\citenamefont
  {Bakhshandeh}, \citenamefont {dos Santos}, \citenamefont {Diehl},\ and\
  \citenamefont {Levin}}]{bakh2015}%
  \BibitemOpen
  \bibfield  {author} {\bibinfo {author} {\bibfnamefont {A.}~\bibnamefont
  {Bakhshandeh}}, \bibinfo {author} {\bibfnamefont {A.~P.}\ \bibnamefont {dos
  Santos}}, \bibinfo {author} {\bibfnamefont {A.}~\bibnamefont {Diehl}}, \ and\
  \bibinfo {author} {\bibfnamefont {Y.}~\bibnamefont {Levin}},\ }\href@noop {}
  {\bibfield  {journal} {\bibinfo  {journal} {The Journal of Chemical Physics}\
  }\textbf {\bibinfo {volume} {142}},\ \bibinfo {pages} {194707} (\bibinfo
  {year} {2015})}\BibitemShut {NoStop}%
\bibitem [{\citenamefont {Bakhshandeh}, \citenamefont {dos Santos},\ and\
  \citenamefont {Levin}(2018)}]{bakh2018}%
  \BibitemOpen
  \bibfield  {author} {\bibinfo {author} {\bibfnamefont {A.}~\bibnamefont
  {Bakhshandeh}}, \bibinfo {author} {\bibfnamefont {A.~P.}\ \bibnamefont {dos
  Santos}}, \ and\ \bibinfo {author} {\bibfnamefont {Y.}~\bibnamefont
  {Levin}},\ }\href@noop {} {\bibfield  {journal} {\bibinfo  {journal} {Soft
  Matter}\ }\textbf {\bibinfo {volume} {14}},\ \bibinfo {pages} {4081}
  (\bibinfo {year} {2018})}\BibitemShut {NoStop}%
\bibitem [{\citenamefont {Bakhshandeh}, \citenamefont {dos Santos},\ and\
  \citenamefont {Levin}(2011)}]{bak2011}%
  \BibitemOpen
  \bibfield  {author} {\bibinfo {author} {\bibfnamefont {A.}~\bibnamefont
  {Bakhshandeh}}, \bibinfo {author} {\bibfnamefont {A.~P.}\ \bibnamefont {dos
  Santos}}, \ and\ \bibinfo {author} {\bibfnamefont {Y.}~\bibnamefont
  {Levin}},\ }\href@noop {} {\bibfield  {journal} {\bibinfo  {journal}
  {Physical review letters}\ }\textbf {\bibinfo {volume} {107}},\ \bibinfo
  {pages} {107801} (\bibinfo {year} {2011})}\BibitemShut {NoStop}%
\bibitem [{\citenamefont {dos Santos}, \citenamefont {Bakhshandeh},\ and\
  \citenamefont {Levin}(2011)}]{dos2011}%
  \BibitemOpen
  \bibfield  {author} {\bibinfo {author} {\bibfnamefont {A.~P.}\ \bibnamefont
  {dos Santos}}, \bibinfo {author} {\bibfnamefont {A.}~\bibnamefont
  {Bakhshandeh}}, \ and\ \bibinfo {author} {\bibfnamefont {Y.}~\bibnamefont
  {Levin}},\ }\href@noop {} {\bibfield  {journal} {\bibinfo  {journal} {The
  Journal of chemical physics}\ }\textbf {\bibinfo {volume} {135}},\ \bibinfo
  {pages} {044124} (\bibinfo {year} {2011})}\BibitemShut {NoStop}%
\bibitem [{\citenamefont {Guerrero-Garc\'{\i}a}\ \emph
  {et~al.}(2005)\citenamefont {Guerrero-Garc\'{\i}a}, \citenamefont
  {Gonz\'alez-Tovar}, \citenamefont {Lozada-Cassou},\ and\ \citenamefont
  {Guevara-Rodr\'iguez}}]{g1}%
  \BibitemOpen
  \bibfield  {author} {\bibinfo {author} {\bibfnamefont {G.~I.}\ \bibnamefont
  {Guerrero-Garc\'{\i}a}}, \bibinfo {author} {\bibfnamefont {E.}~\bibnamefont
  {Gonz\'alez-Tovar}}, \bibinfo {author} {\bibfnamefont {M.}~\bibnamefont
  {Lozada-Cassou}}, \ and\ \bibinfo {author} {\bibfnamefont {F.~J.}\
  \bibnamefont {Guevara-Rodr\'iguez}},\ }\href {\doibase 10.1063/1.1949168}
  {\bibfield  {journal} {\bibinfo  {journal} {The Journal of Chemical Physics}\
  }\textbf {\bibinfo {volume} {123}},\ \bibinfo {pages} {034703} (\bibinfo
  {year} {2005})},\ \Eprint
  {http://arxiv.org/abs/https://doi.org/10.1063/1.1949168}
  {https://doi.org/10.1063/1.1949168} \BibitemShut {NoStop}%
\bibitem [{\citenamefont {Guerrero-Garc\'{\i}a}, \citenamefont
  {Gonz\'alez-Tovar},\ and\ \citenamefont {ChMavez-P\'aez}(2009)}]{g2}%
  \BibitemOpen
  \bibfield  {author} {\bibinfo {author} {\bibfnamefont {G.~I.}\ \bibnamefont
  {Guerrero-Garc\'{\i}a}}, \bibinfo {author} {\bibfnamefont {E.}~\bibnamefont
  {Gonz\'alez-Tovar}}, \ and\ \bibinfo {author} {\bibfnamefont
  {M.}~\bibnamefont {ChMavez-P\'aez}},\ }\href {\doibase
  10.1103/PhysRevE.80.021501} {\bibfield  {journal} {\bibinfo  {journal} {Phys.
  Rev. E}\ }\textbf {\bibinfo {volume} {80}},\ \bibinfo {pages} {021501}
  (\bibinfo {year} {2009})}\BibitemShut {NoStop}%
\bibitem [{\citenamefont {Guerrero-Garc\'{\i}a}\ \emph
  {et~al.}(2016)\citenamefont {Guerrero-Garc\'{\i}a}, \citenamefont
  {Gonz\'alez-Tovar}, \citenamefont {Quesada-Perez},\ and\ \citenamefont
  {Martin-Molina}}]{g3}%
  \BibitemOpen
  \bibfield  {author} {\bibinfo {author} {\bibfnamefont {G.~I.}\ \bibnamefont
  {Guerrero-Garc\'{\i}a}}, \bibinfo {author} {\bibfnamefont {E.}~\bibnamefont
  {Gonz\'alez-Tovar}}, \bibinfo {author} {\bibfnamefont {M.}~\bibnamefont
  {Quesada-Perez}}, \ and\ \bibinfo {author} {\bibfnamefont {A.}~\bibnamefont
  {Martin-Molina}},\ }\href {\doibase 10.1039/C6CP03483G} {\bibfield  {journal}
  {\bibinfo  {journal} {Phys. Chem. Chem. Phys.}\ }\textbf {\bibinfo {volume}
  {18}},\ \bibinfo {pages} {21852} (\bibinfo {year} {2016})}\BibitemShut
  {NoStop}%
\bibitem [{\citenamefont {Levin}, \citenamefont {Barbosa},\ and\ \citenamefont
  {Tamashiro}(1998)}]{marcia}%
  \BibitemOpen
  \bibfield  {author} {\bibinfo {author} {\bibfnamefont {Y.}~\bibnamefont
  {Levin}}, \bibinfo {author} {\bibfnamefont {M.~C.}\ \bibnamefont {Barbosa}},
  \ and\ \bibinfo {author} {\bibfnamefont {M.~N.}\ \bibnamefont {Tamashiro}},\
  }\href@noop {} {\bibfield  {journal} {\bibinfo  {journal} {Europhysics
  Letters}\ }\textbf {\bibinfo {volume} {41}},\ \bibinfo {pages} {123}
  (\bibinfo {year} {1998})}\BibitemShut {NoStop}%
\bibitem [{\citenamefont {Naji}\ \emph {et~al.}(2014)\citenamefont {Naji},
  \citenamefont {Ghodrat}, \citenamefont {Komaie-Moghaddam},\ and\
  \citenamefont {Podgornik}}]{naji}%
  \BibitemOpen
  \bibfield  {author} {\bibinfo {author} {\bibfnamefont {A.}~\bibnamefont
  {Naji}}, \bibinfo {author} {\bibfnamefont {M.}~\bibnamefont {Ghodrat}},
  \bibinfo {author} {\bibfnamefont {H.}~\bibnamefont {Komaie-Moghaddam}}, \
  and\ \bibinfo {author} {\bibfnamefont {R.}~\bibnamefont {Podgornik}},\ }\href
  {\doibase 10.1063/1.4898663} {\bibfield  {journal} {\bibinfo  {journal} {The
  Journal of Chemical Physics}\ }\textbf {\bibinfo {volume} {141}},\ \bibinfo
  {pages} {174704} (\bibinfo {year} {2014})},\ \Eprint
  {http://arxiv.org/abs/https://doi.org/10.1063/1.4898663}
  {https://doi.org/10.1063/1.4898663} \BibitemShut {NoStop}%
\bibitem [{\citenamefont {Palaia}\ \emph {et~al.}(2018)\citenamefont {Palaia},
  \citenamefont {Trulsson}, \citenamefont {Šamaj},\ and\ \citenamefont
  {Trizac}}]{Trizac}%
  \BibitemOpen
  \bibfield  {author} {\bibinfo {author} {\bibfnamefont {I.}~\bibnamefont
  {Palaia}}, \bibinfo {author} {\bibfnamefont {M.}~\bibnamefont {Trulsson}},
  \bibinfo {author} {\bibfnamefont {L.}~\bibnamefont {Šamaj}}, \ and\ \bibinfo
  {author} {\bibfnamefont {E.}~\bibnamefont {Trizac}},\ }\href {\doibase
  10.1080/00268976.2018.1471234} {\bibfield  {journal} {\bibinfo  {journal}
  {Molecular Physics}\ }\textbf {\bibinfo {volume} {0}},\ \bibinfo {pages} {1}
  (\bibinfo {year} {2018})},\ \Eprint
  {http://arxiv.org/abs/https://doi.org/10.1080/00268976.2018.1471234}
  {https://doi.org/10.1080/00268976.2018.1471234} \BibitemShut {NoStop}%
\bibitem [{\citenamefont {Messina}(2002)}]{messina}%
  \BibitemOpen
  \bibfield  {author} {\bibinfo {author} {\bibfnamefont {R.}~\bibnamefont
  {Messina}},\ }\href@noop {} {\bibfield  {journal} {\bibinfo  {journal} {The
  Journal of chemical physics}\ }\textbf {\bibinfo {volume} {117}},\ \bibinfo
  {pages} {11062} (\bibinfo {year} {2002})}\BibitemShut {NoStop}%
\bibitem [{\citenamefont {{J D Jackson}}(1999)}]{Jackson}%
  \BibitemOpen
  \bibfield  {author} {\bibinfo {author} {\bibnamefont {{J D Jackson}}},\
  }\href@noop {} {\emph {\bibinfo {title} {Classical Electrodynamics}}}\
  (\bibinfo  {publisher} {Wiley},\ \bibinfo {address} {New York},\ \bibinfo
  {year} {1999})\BibitemShut {NoStop}%
\bibitem [{\citenamefont {dos Santos}\ and\ \citenamefont
  {Levin}(2018)}]{dos20181}%
  \BibitemOpen
  \bibfield  {author} {\bibinfo {author} {\bibfnamefont {A.~P.}\ \bibnamefont
  {dos Santos}}\ and\ \bibinfo {author} {\bibfnamefont {Y.}~\bibnamefont
  {Levin}},\ }\href@noop {} {\bibfield  {journal} {\bibinfo  {journal} {The
  Journal of Chemical Physics}\ }\textbf {\bibinfo {volume} {148}},\ \bibinfo
  {pages} {222817} (\bibinfo {year} {2018})}\BibitemShut {NoStop}%
\bibitem [{\citenamefont {Adar}, \citenamefont {Markovich},\ and\ \citenamefont
  {Andelman}(2017)}]{adar2017bjerrum}%
  \BibitemOpen
  \bibfield  {author} {\bibinfo {author} {\bibfnamefont {R.~M.}\ \bibnamefont
  {Adar}}, \bibinfo {author} {\bibfnamefont {T.}~\bibnamefont {Markovich}}, \
  and\ \bibinfo {author} {\bibfnamefont {D.}~\bibnamefont {Andelman}},\
  }\href@noop {} {\bibfield  {journal} {\bibinfo  {journal} {The Journal of
  Chemical Physics}\ }\textbf {\bibinfo {volume} {146}},\ \bibinfo {pages}
  {194904} (\bibinfo {year} {2017})}\BibitemShut {NoStop}%
\bibitem [{\citenamefont {Levin}\ and\ \citenamefont
  {Flores-Mena}(2001)}]{levin21}%
  \BibitemOpen
  \bibfield  {author} {\bibinfo {author} {\bibfnamefont {Y.}~\bibnamefont
  {Levin}}\ and\ \bibinfo {author} {\bibfnamefont {J.~E.}\ \bibnamefont
  {Flores-Mena}},\ }\href@noop {} {\bibfield  {journal} {\bibinfo  {journal}
  {EPL (Europhysics Letters)}\ }\textbf {\bibinfo {volume} {56}},\ \bibinfo
  {pages} {187} (\bibinfo {year} {2001})}\BibitemShut {NoStop}%
\bibitem [{\citenamefont {Jadhao}, \citenamefont {Solis},\ and\ \citenamefont
  {de~la Cruz}(2012)}]{monica}%
  \BibitemOpen
  \bibfield  {author} {\bibinfo {author} {\bibfnamefont {V.}~\bibnamefont
  {Jadhao}}, \bibinfo {author} {\bibfnamefont {F.~J.}\ \bibnamefont {Solis}}, \
  and\ \bibinfo {author} {\bibfnamefont {M.~O.}\ \bibnamefont {de~la Cruz}},\
  }\href {\doibase 10.1103/PhysRevLett.109.223905} {\bibfield  {journal}
  {\bibinfo  {journal} {Phys. Rev. Lett.}\ }\textbf {\bibinfo {volume} {109}},\
  \bibinfo {pages} {223905} (\bibinfo {year} {2012})}\BibitemShut {NoStop}%
\bibitem [{\citenamefont {y~Wang}\ and\ \citenamefont {q~Ma}(2012)}]{Wang}%
  \BibitemOpen
  \bibfield  {author} {\bibinfo {author} {\bibfnamefont {Z.}~\bibnamefont
  {y~Wang}}\ and\ \bibinfo {author} {\bibfnamefont {Y.}~\bibnamefont {q~Ma}},\
  }\href {\doibase 10.1103/PhysRevE.85.062501} {\bibfield  {journal} {\bibinfo
  {journal} {Phys. Rev. E}\ }\textbf {\bibinfo {volume} {85}},\ \bibinfo
  {pages} {062501} (\bibinfo {year} {2012})}\BibitemShut {NoStop}%
\bibitem [{\citenamefont {Wang}(2016)}]{wang1}%
  \BibitemOpen
  \bibfield  {author} {\bibinfo {author} {\bibfnamefont {Z.-Y.}\ \bibnamefont
  {Wang}},\ }\href {http://stacks.iop.org/1742-5468/2016/i=4/a=043205}
  {\bibfield  {journal} {\bibinfo  {journal} {Journal of Statistical Mechanics:
  Theory and Experiment}\ }\textbf {\bibinfo {volume} {2016}},\ \bibinfo
  {pages} {043205} (\bibinfo {year} {2016})}\BibitemShut {NoStop}%
\bibitem [{\citenamefont {Wang}, \citenamefont {Zhang},\ and\ \citenamefont
  {Ma}(2018)}]{Wang2}%
  \BibitemOpen
  \bibfield  {author} {\bibinfo {author} {\bibfnamefont {Z.-Y.}\ \bibnamefont
  {Wang}}, \bibinfo {author} {\bibfnamefont {P.}~\bibnamefont {Zhang}}, \ and\
  \bibinfo {author} {\bibfnamefont {Z.}~\bibnamefont {Ma}},\ }\href@noop {}
  {\bibfield  {journal} {\bibinfo  {journal} {Phys. Chem. Chem. Phys.}\
  }\textbf {\bibinfo {volume} {20}},\ \bibinfo {pages} {4118} (\bibinfo {year}
  {2018})}\BibitemShut {NoStop}%
\bibitem [{\citenamefont {Guerrero~García}\ and\ \citenamefont {Olvera de~la
  Cruz}(2014)}]{gl}%
  \BibitemOpen
  \bibfield  {author} {\bibinfo {author} {\bibfnamefont {G.~I.}\ \bibnamefont
  {Guerrero~García}}\ and\ \bibinfo {author} {\bibfnamefont {M.}~\bibnamefont
  {Olvera de~la Cruz}},\ }\href {\doibase 10.1021/jp5045173} {\bibfield
  {journal} {\bibinfo  {journal} {The Journal of Physical Chemistry B}\
  }\textbf {\bibinfo {volume} {118}},\ \bibinfo {pages} {8854} (\bibinfo {year}
  {2014})},\ \bibinfo {note} {pMID: 24953671},\ \Eprint
  {http://arxiv.org/abs/https://doi.org/10.1021/jp5045173}
  {https://doi.org/10.1021/jp5045173} \BibitemShut {NoStop}%
\bibitem [{\citenamefont {Lindell}()}]{Lindell}%
  \BibitemOpen
  \bibfield  {author} {\bibinfo {author} {\bibfnamefont {I.~V.}\ \bibnamefont
  {Lindell}},\ }\href {\doibase 10.1029/91RS02255} {\bibfield  {journal}
  {\bibinfo  {journal} {Radio Science}\ }\textbf {\bibinfo {volume} {27}},\
  \bibinfo {pages} {1}},\ \Eprint
  {http://arxiv.org/abs/https://agupubs.onlinelibrary.wiley.com/doi/pdf/10.1029/91RS02255}
  {https://agupubs.onlinelibrary.wiley.com/doi/pdf/10.1029/91RS02255}
  \BibitemShut {NoStop}%
\bibitem [{\citenamefont {Norris}(1995)}]{Norris}%
  \BibitemOpen
  \bibfield  {author} {\bibinfo {author} {\bibfnamefont {W.~T.}\ \bibnamefont
  {Norris}},\ }\href
  {http://digital-library.theiet.org/content/journals/10.1049/ip-smt_19951564}
  {\bibfield  {journal} {\bibinfo  {journal} {IEE Proceedings - Science,
  Measurement and Technology}\ }\textbf {\bibinfo {volume} {142}},\ \bibinfo
  {pages} {142} (\bibinfo {year} {1995})}\BibitemShut {NoStop}%
\bibitem [{\citenamefont {{D Frenkel and B Smith}}(1996)}]{Frenkel}%
  \BibitemOpen
  \bibfield  {author} {\bibinfo {author} {\bibnamefont {{D Frenkel and B
  Smith}}},\ }\href@noop {} {\emph {\bibinfo {title} {Understanding Molecular
  Simulation}}}\ (\bibinfo  {publisher} {Academic Press},\ \bibinfo {address}
  {New York},\ \bibinfo {year} {1996})\BibitemShut {NoStop}%
\bibitem [{\citenamefont {{M P Allen and D J Tildesley}}(1987)}]{Allen}%
  \BibitemOpen
  \bibfield  {author} {\bibinfo {author} {\bibnamefont {{M P Allen and D J
  Tildesley}}},\ }\href@noop {} {\emph {\bibinfo {title} {Computer Simulations
  of Liquids}}}\ (\bibinfo  {publisher} {Oxford University Press},\ \bibinfo
  {address} {Oxford},\ \bibinfo {year} {1987})\BibitemShut {NoStop}%
\end{thebibliography}%
\end{document}